\documentclass[10.5pt]{report}
\usepackage{latexsym}
\usepackage{amsmath}
\pdfoutput=1
\usepackage{times}
\usepackage{amssymb}
\usepackage{fancyheadings}
\usepackage[T1]{fontenc}
\usepackage[utf8]{inputenc}
\usepackage{url}
\usepackage{hyperref}
\usepackage{color}
\usepackage{graphicx}
 \usepackage{epsfig}

\date{}

\begin{document}

\title{\bf Functional Geometry of Human Connectome and  Robustness of
  Gender Differences}
\author{Bosiljka Tadi\'c$^{a,b}$, Miroslav Andjelkovi\'c$^c$, Roderick Melnik$^d$\\
$^a$Department of Theoretical Physics, Jo\v zef Stefan Institute,
Jamova 39, Ljubljana, Slovenia; \\$^b$Complexity Science Hub,
Josefstaedter Strasse 39, Vienna, Austria;\\
$^c$ Institute for Nuclear Sciences Vin\v ca,  University of Belgrade,
11000 Belgrade, Serbia;\\ $^d$MS2Discovery Interdisciplinary Research
Institute, M2NeT Laboratory and Department of Mathematics,\\ Wilfrid
Laurier University, Waterloo, ON, Canada}
\maketitle
\begin{abstract}
\noindent 
Mapping the brain imaging data to networks, where each node represents a specific area of the brain, has enabled an objective graph-theoretic analysis of human connectome. However, the latent structure on higher-order connections remains unexplored, where many brain regions acting in synergy perform complex functions.
Here we analyse this hidden structure using the simplicial complexes parametrisation where the shared faces of simplexes encode higher-order relationships between groups of nodes and emerging hyperbolic geometry. 
Based on data collected within the Human Connectome Project, we perform a systematic analysis of consensus networks of 100 female (F-connectome) and 100 male (M-connectome) subjects by varying the number of fibres launched. 
Our analysis reveals that the functional geometry of the common F\&M-connectome coincides with the M-connectome and is characterized by a complex architecture of simplexes to the 14th order, which is built in six anatomical communities, and short cycles among them. Furthermore,  the F-connectome has additional connections that involve different brain regions, thereby increasing the size of simplexes and introducing new cycles. 
By providing new insights into the internal organisation of  anatomical brain modules as well as into the links  between them that are essential to dynamics, these results  also highlight the functional gender-related differences.
\\[3pt]
\end{abstract}

\flushbottom
\maketitle

\thispagestyle{empty}

\section*{Introduction\label{sec-intro}}
Human psychology and behaviour are determined by functional brain
connectivity among neurons, neural assemblies, or entire regions, making the patterns of circuitry that can be
detected by brain imaging  \cite{BrainImaging_Babiloni2005}. Recent large-scale research into the brain imaging data within the Human Connectome Project (HCP) \cite{HCP_Neuroimage,ganepola2017,CParcellation_2018} aims to uncover, describe and understand the functional structure of human connectome; the connectome is visualised as a network consisting of different brain regions  (grey matter) and paths between them (white-matter fibre bundles) that can be determined by mapping the diffusion-MRI and tractography data.  The network nodes are identified as distinct brain regions that are functionally similar and spatially close as well as equally connected to the other regions
\cite{BrainConnectivityNets_2010,WholeBrainNet-Nodes_Zalesky2010,BrainNetParcellation_Shen2010,CParcellation_2018}. The connections between these regions, which are determined from brain imaging data, can depend on a number of factors, and vary among different subjects, performed tasks and conditions.  Therefore, the consensus between the pipelines in the structural connectome can be mapped from a large population tractography data
\cite{Consensus_AnatomicalBrainPipelines_Plos2014} and depends on many parameters. 
Based on the data from HCP \cite{HCP_Neuroimage} and the brain mapper developed in \cite{BrainMapping_Zhang2018}, the Budapest connectome server  \cite{http-Budapest-server3.0} provides the possibilities to infer the \textit{consensus networks} at a variety of the relevant parameters, as described in
\cite{Hung1,Hung2}. The mapping of imaging data to the brain
networks enables an objective analysis based on graph theory methods \cite{Sporns_BrainNets_review2013,brainGraph_ideasSpanci2014}.

Recently, different studies of brain imaging data revealed the strong  evidences for gender differences in the structural connectome
\cite{brainGender_minireview2017,brainGender_meta2014,GD_cognitiveSci2014,genderData_age2011,functionGender_HBM,brainGender_perfusionPLOS2015,GD_Bud_PLOS2015}. This
subject was not well researched, but already it brought some controversial debates \cite{GD_PNAS2014}. 
The exact origin of these gender differences and their
potentials and impact on the level of individual and social behaviour
are still to be investigated \cite{GD_ageMetabolicPNAS2019}. On the other hand, the current degree of reliability of the connectome data provides an opportunity for a mathematical analysis of structural differences at all levels. 
For example, a recent study \cite{GD_Bud_PLOS2015}  has shown that the consensus female connectome has superior connectivity than the consensus male connectome in many graph-theoretic measures. 

Recent investigations of geometrical properties of various complex systems
\cite{Geometry_NegC_biol,Geometry_NegC_traffic,we-PhysA2015,we-MBNets-PLOS2016,hidden-multiplexNat2016,Homology_SalnikovEJP2018,we-Frontiers2018,Geometry_BianconiSR2017,we_SA2017,Geometry_weEJP2018} show the relevance of the higher-order connectivity beyond standardly considered  pairways interactions. 
Mathematically, the impact of these higher order interactions is adequately described by the simplicial complexes in the algebraic topology of graphs \cite{jj-book,kozlov-book,cliquecomplexes,SM-book2017}. 
In these complexes, elementary geometrical shapes (triangles, tetrahedra, and simplexes of higher order) are combined through shared substructures of various orders. These geometrical structures directly influence dynamic processes that the complex system in question performs, such as transport, diffusion, or synchronisation among the involved nodes.
In the case of brain networks, the main dynamic function pertains  to
maintaining an optimal balance between the processes of  integration and
segregation where different regions of the brain can be simultaneously
involved and the present modular structure of
the brain plays an important role \cite{RevModPhys2018,NatRevNeuro_segregIntegration2015,brainNets_tasksPNAS2015,Dyn-complexity-SciRep2016}. Anatomical modules
of the brain, which are recognized as different mesoscopic communities in the
network \cite{brainCommunities_FrontNeurosci2010,franco2014,brain-mesoscopic1,brain-mesoscopic2}, are based on spatial topography and coexpression of genes in the brain cells \cite{brainNets_modulesTranscriptomeNatneurosci2008}. 
It has been suggested that each module performs a discrete cognitive
function while specific connector nodes take on communication between
modules \cite{brainNets_tasksPNAS2015}.
However, the fine functional organisation inside these modules remains unexplored.
Besides, the occurrence of simplicial complexes causes the emergent hyperbolicity or a negative curvature \cite{Gromov1987} in the structure of the graph, which affects its functional properties.
In this sense, the complete graph and associated tree are ideally hyperbolic,
characterised by the hyperbolicity parameter  $\delta=0$. The graphs
with small values of  $\delta$ are subject to intensive investigations
for their ubiquity in natural and social systems, as well as in technology applications
\cite{HB-Bermudo2016,Geometry_NegC_biol,Geometry_NegC_traffic,we-Frontiers2018,Geometry_weEJP2018}.
 Moreover, current theoretical studies reveal that Gromov hyperbolic graphs with a small hyperbolicity parameter have specific mathematical properties \cite{HB-Bermudo2016}. In particular, the bounds for the $\delta$-parameter of the whole graph can be derived  from subjacent simpler graphs,
for example, induced cycles or clique separators of a given length \cite{HB-Chepoi2008-D-centers,HB-BermudoHBviasmallerGraph2013,HB-distortionDelta2014,Hyperbolicity_cliqueDecomposition2017,Hyperbolicity_chordality2011,Hyperbolicity_chordality2017}.
Therefore, the study of the hyperbolicity of brain graphs can reveal
the presence of typical local structures that are potentially
decomposable into some known forms, which underlie the brain's dynamic complexity.

In this work, we considerably expand the analysis of human
connectome beyond the simple pairwise connectivity. Using the
mathematical techniques of algebraic topology of graphs, we identify hierarchically organised complexes
that encode higher-order relationships between regions of the brain and explore the hyperbolic geometry of brain graphs. 
We consider the consensus connectomes mapped from 100 female (F-connectome) and 100  male (M-connectome)
subjects, using the brain mapper and imaging data from the Human Connectome Project, which is provided by the Budapest server 3.0 \cite{http-Budapest-server3.0}.  
The weighted edges are inferred according to the \textit{electrical connectivity} criteria, which are most sensitive to the
number of fibres observed in the tractography data. We analyse the connectomes that correspond to the significant variation in the number of fibres launched (see Methods). 
With the appropriate topology measures, our objectives are to determine the hidden structure of human connectome endowed with the relationships between groups of nodes and express the possible gender differences in this context.
 To this end, we construct and investigate a common F\&M-connectome at  different numbers of fibres and determine its structure, parametrised by simplicial complexes, and the graph's hyperbolicity parameter. Furthermore, by comparing edges in the F- and M-connectomes,  we identify the excess edges that appear consistently in the F-connectome with an increased number of
fibres. Our mathematical analysis reveals a rich structure of
simplicial complexes that are common to the F\&M-connectome and belong
to different brain anatomical communities and cycles that connect them
inside and across the two brain hemispheres. It further confirms the
higher connectivity of the F-connectome and demonstrates that  the excess edges have a well-organised structure that includes a particular set of paths and brain regions.

\section*{Methods\label{sec:methods}}

\textbf{Input Data \& Consensus Networks.}
We downloaded the data for male and female connectomes from the Budapest connectome server 3.0 \cite{http-Budapest-server3.0}. Using the data from HCP \cite{HCP_Neuroimage} and the brain mapper provided in \cite{BrainMapping_Zhang2018} the server produces the connectome corresponding to
the settings of a variety of parameters, the meaning of which is specified in \cite{Hung1,Hung2}.
For our study, we have selected the data that provide the
\textit{consensus networks} for female connectome and male connectome based on 100 subjects of each gender. The corresponding brain networks consist of N=1015 nodes (brain regions) and the weights of the connections between them determined according to the \textit{electrical connectivity} criteria, i.e., the number of fibres between the considered pair of regions is divided by the average fibre length. We consider three different fibre counts, comprising of $N_F=$ 20K, 200K, and 1000K fibres, where for short K$\equiv$ 1000. For the
additional parameters, we have set the minimum edge confidence as 100\%, minimum edge weight as 4, and the median weight calculation. The resulting adjacency matrices of the weighted networks, herewith called  F-connectome and M-connectome, respectively, are downloaded together with the node labels, i.e., standardly accepted names of the brain regions.

\textbf{Gromov hyperbolicity parameter of graphs.}
A generalization of the
Gromov notion of hyperbolicity \cite{Gromov1987} is applied to graphs
endowed with the shortest-path metric. Specifically, the 4-point Gromov criterion states that a
graph $G$ is $\delta$-hyperbolic $iff$ for any four vertices
$(A,B,C,D)$ there is a fixed small value $\delta(G)$ such that the following relation beween the sums of distances
${\cal{S}}\equiv d(A,B)+d(C,D) \leq {\cal{M}}\equiv  d(A,C) + d(B,D)
\leq {\cal{L}}\equiv d(A,D)+ d(B,C)$
implies $d(A,D)+ d(B,C) - d(A,C) - d(B,D)  \leq 2\delta(G)$.
Thus, for a $\delta$-hyperbolic graph, there is  $\delta (G)$ such that any four nodes of the graph satisfy the condition
\begin{equation}
\delta(A,B,C,D) \equiv  \frac{{\cal{L}} - {\cal{M}}}{2} \leq \delta (G)\; .
\label{eq:hb-condition}
\end{equation}
From the triangle inequality, the value of
 $({\cal{L}}-{\cal{M}})/2$ is bounded brom above by the minimal distance
 $d_{min}\equiv min\{d(A,B),d(C,D)\} $ in the smallest sum
 ${\cal{S}}$.
This relationship enables a  direct computation of the hyperbolicity
parameter of a graph, which is given by its adjacency matrix. In particular,  by
sampling a large number ($10^9$)  4-tuples of vertices we plot  $\delta(A,B,C,D) $
against the corresponding $d_{min}$; the plot  saturates at larger
distances.   We compute the average   $\langle \delta \rangle$ for all $d_{min}$ as well as
$\delta_{max}=max_G\{\delta(A,B,C,D)\}$, which gives $\delta (G)$. \\
We also determine  the distribution $P(d)$ of the shortest-path distances
$d$ on the graph. The largest distance defines the
graph's \textit{diameter} $D$, which gives the upper bound to the hyperbolicity parameter, $\delta(G)\leq
D/2$. As mentioned above, the hyperbolic graphs with a small parameter
$\delta$  have a specific structure of subgraphs, from which the upper
bound of $\delta(G)$ can be derived \cite{HB-BermudoHBviasmallerGraph2013,HB-distortionDelta2014,Hyperbolicity_cliqueDecomposition2017,Hyperbolicity_chordality2011,Hyperbolicity_chordality2017}.
In this context, the following definitions apply. A subgraph $\Gamma$ of $G$ is called \textit{isometric} if the
distance between every pair of vertices $(A,B)\in \Gamma$ is equal to
the distance between them measured on $G$, i.e.,
$d_\Gamma(A,B)=d_G(A,B)$. A \textit{cycle} $C_n$ is a sequence of $n$
pairwise connected vertices with $n+1\to 1$; an \textit{induced cycle}
does not contain a \textit{chord}, an edge connecting nonconsecutive
vertices. A \textit{clique} of size $s\equiv q_{max}+1$ is the full graph of $s$
vertices and $s(s-1)/2$ edges.

\textbf{Q-analysis of graphs: definition of structure
  vectors.}
Considering a connectome as an undirected and unweighted graph $G$, the
higher-order connectivity of its vertices can be appropriately parametrised  by the maximal complete subgraphs (or cliques) whose vertices belong
to a clique complex $C(G)$ in the graph $G$ \cite{cliquecomplexes}. Two cliques $\sigma_r$
and $\sigma _q$ of the orders $r,q$ can be interconnected by sharing
some vertices; then the structure made by the shared vertices represents a
common face of both cliques. For example, if for $r<q$ all vertices of $\sigma_r$ belong to
$\sigma_q$, then the simplex $\sigma_r$ represents a face of the
order $r$ in the simplex
$\sigma_q$.  The simplicial complex represents the aggregate of
cliques  that share the faces of different orders
$q=0,1,2\cdots q_{max}^\prime-1$, where  $q_{max}^\prime$ indicates the order of the
largest clique in the complex.  
The order of a  simplicial complex is the largest order of a simplex
in it;  we denote by $q_{max}$ the order of the largest complex in
the entire graph.

Applying  the Bron-Kerbosch algorithm \cite{Bron}, the adjacency matrix
of the graph $G$ is converted into the incidence matrix $\Lambda$,
which contains all cliques in the graph by identifying the vertices that belong to them; using this
information, we then find
how different cliques interconnect via shared nodes to make the
higher-order  structures. 
The  overall hierarchical organisation of the graph can be quantified 
\cite{Qanalysis1,Qanalysis2,ATKIN1976,MilanLNCS,we-PhysA2015,we-MBNets-PLOS2016}
by three \textit{structure
  vectors} having the components along different topology levels
$q=1,2,3,\cdots q_{max}$. Specifically, for each considered graph, we
determine:
\begin{itemize}
\item FSV---the first structure vector $\{Q_q\}$, where each
  $Q_q$ represents the number of $q$-connected components;
\item SSV---the second structure vector  $\{n_q\}$, where $n_q$
  indicates the number of connected  components from the level $q$ upwards;
\item TSV---the third structure vector $\{{\hat{Q}}_q\}$ is introduced
  to quantify the degree of interconnectivity between cliques at each level
  $q$, and can be derived from the other two as $\{{\hat{Q}}_q\}=1-Q_q/n_q$.
\end{itemize}
These structure vectors  provide a measure of the graph's  global
architecture (see \cite{we-PRE} for the application of $Q$-analysis
for the vertex neighbourhood). For completeness, we also determine standard
graph measures \cite{bb-graphtheory,sergey-lectures}, and community
structure \cite{franco2014,mitrovic2009,lancichinetti2010} of the
typical connectome graph, see Results. 
\textit{Visualisation and standard graph parameters} are made by using Gephi software \cite{Gephi-description}.

\section*{Results\label{sec:results}}
\subsection*{Consensus Networks of Human Connectome}
\begin{figure}[!htb]
\begin{tabular}{cc} 
\resizebox{16pc}{!}{\includegraphics{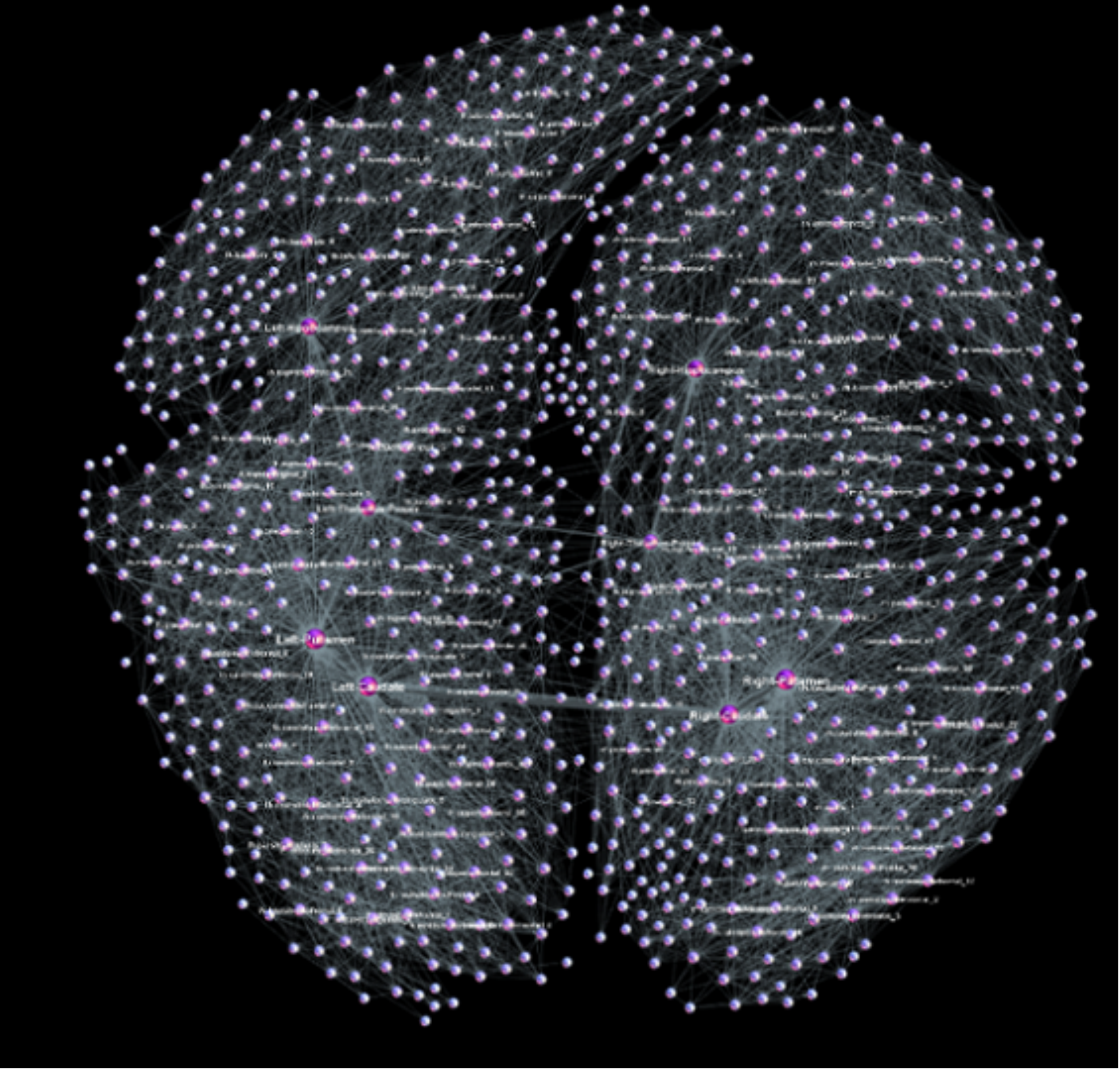}}\\
\end{tabular}
\caption{The female connectome at the highest resolution consisting of 1115 nodes (brain regions) and 11339 edges between them.
  The network is deduced from the HCP data provided at the server \cite{http-Budapest-server3.0} with weighted edges as the median for 100 female subjects and $N_F=$1000K fibres launched between each pair of nodes. 
}
\label{fig-Fconn1000K}
\end{figure}

According to the parameter settings (see Methods), the considered
F-connectome consists of the edges that appear in all 100 female
subjects, and similarly, the M-connectome contains the edges that are present in all 100 male subjects. For the illustration, the F-connectome at 1000K fibres is shown in Fig.\ \ref{fig-Fconn1000K} with the labelled brain regions as nodes.
Here, we use the simplicial complexes parametrisation (see Methods) and the graph's hyperbolicity measures to uncover the hidden structure of human connectome, which is encoded in the higher-order connectivity between groups of nodes. Furthermore, using these mathematical measures, we analyse the variations of the brain connectivity patterns depending on the number of fibres launched and the gender of the subjects. As we will show in the following, these differ significantly depending on the number of launched fibres $N_F$ and between the genders. 

\begin{figure}[!htb]
\begin{tabular}{cc} 
\resizebox{14pc}{!}{\includegraphics{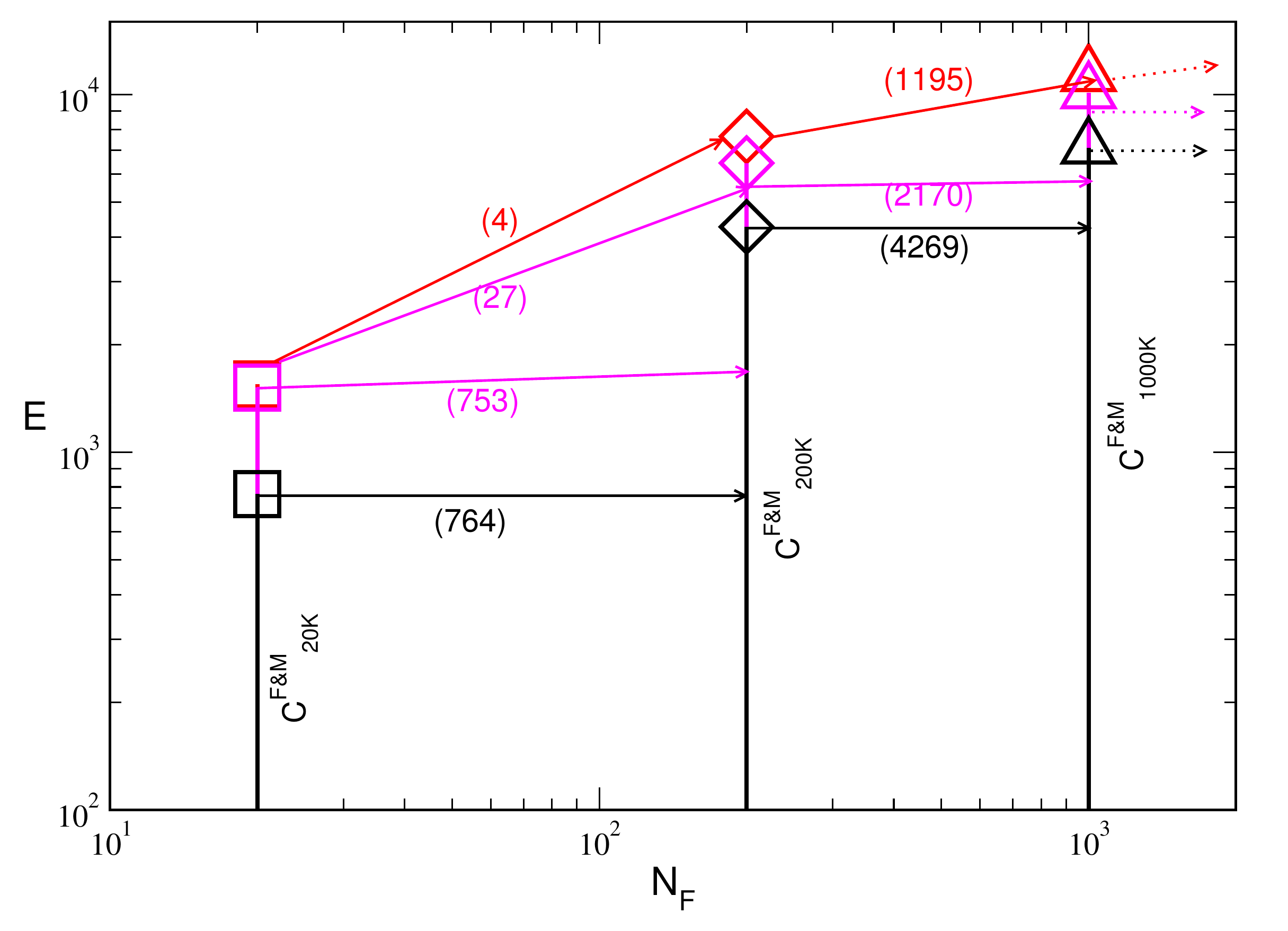}}\\
\end{tabular}
\caption{Schematic view of the number of edges $E$ and their co-occurrence in the connectomes at the increasing number of fibres $N_F$, see also Table\ \ref{tab-EvsNf}. The common $C^{F\&M}$-connectome at a large number of fibres, $N_F$, inherits all edges from the $C^{F\&M}$ at a lower $N_F$, black lines, and a fraction of the excess edges of the F-connectome, shown by pink lines. The top line (red) shows the number of robust excess edges in F-connectome which do not appear in any of the larger common $C^{F\&M}$-connectomes. }
\label{fig-shema}
\end{figure}

\begin{table}[!h]
\caption{For the number of launched fibres $N_F$, the corresponding number of edges are shown in the consensus male (M) and female (F) connectomes, the  edges $C^{F\&M}$  common to F\&M connectomes, and the total number $F^{e0}$ of  excess edges in the F-connectome; the fractions of $F^{e0}$ indicated as  $F^{ec+}$ and  $F^{ecc+}$ are the edges that appear in the common connectomes at the two higher $N_F$, respectively, while $F^{ex}$ are the excess edges also at the higher $N_F$. The difference between $M$ and $C^{F\&M}$  at 20K and 200K consists of 12 and 16 edges, which all appear in
$C^{F\&M}$ at 1000K.}
\label{tab-EvsNf}
\begin{tabular}{|c|cc|c|cccc|}
\hline
$N_F$& M& F& $C^{F\&M}$& $F^{e0}$& $F^{ec+}$&$F^{ecc+}$&$F^{ex}$\\
\hline
20000& 776&1548& 764& 784& 753&27&4\\
200000&4285&7634& 4269& 3365& 2170&-& 1195\\
1000000& 7110&11339& 7110& 4229& -&-&$\geq$1195\\
\hline
\end{tabular}
\end{table}

\begin{figure*}[!htb]
\begin{tabular}{cc} 
\resizebox{24pc}{!}{\includegraphics{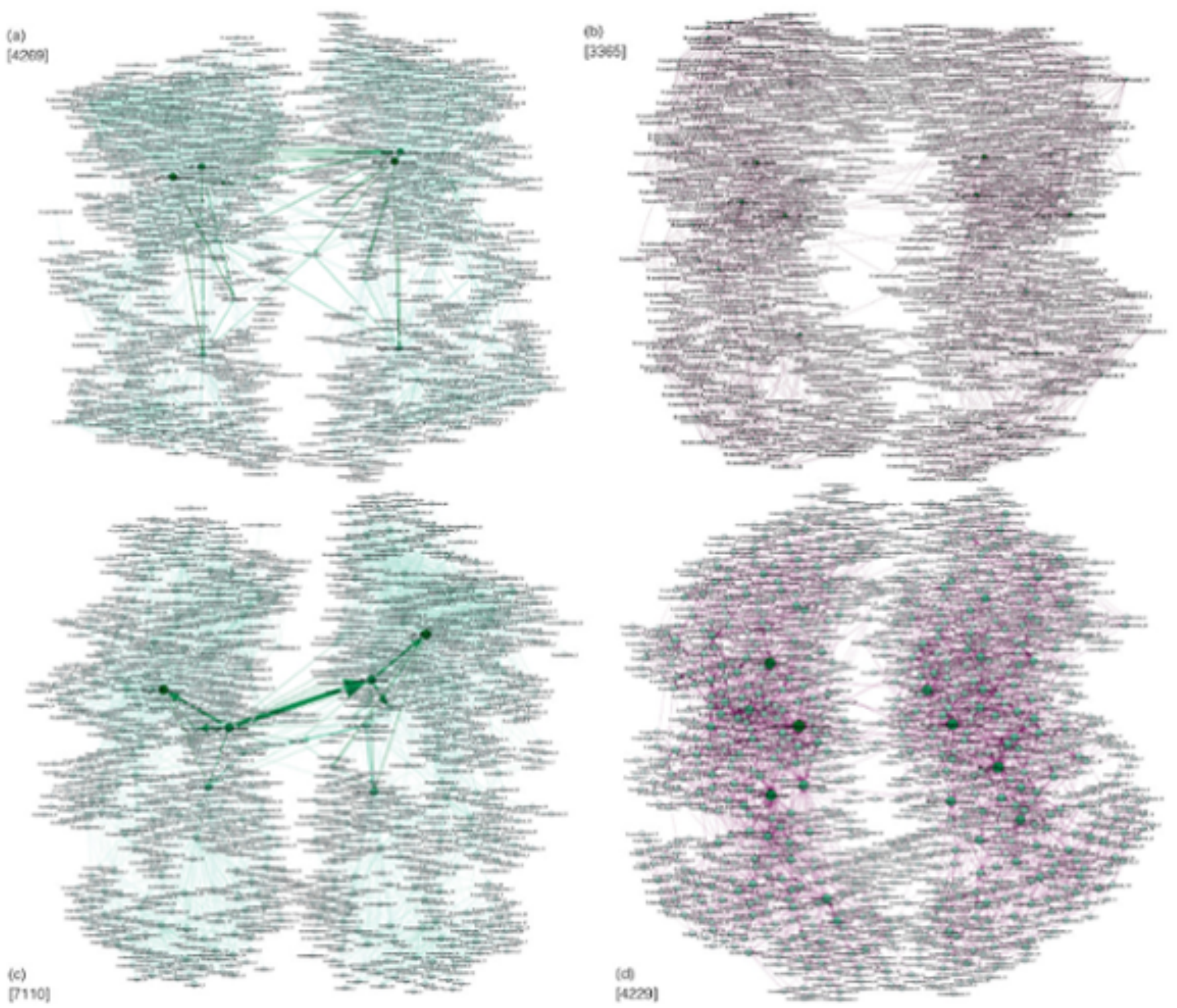}}
\end{tabular}
\caption{Networks of connections established among labelled brain
  regions at  different numbers of launched fibres $N_F$:  (a) Common
  M\&F connectome at $N_F=200K$ and (c) common M\&F connectome at
  $N_F=1000K$, the weights of M-connectome are shown. (b) and (d) The
  patterns of the additional edges appearing in the F-connectome
  (F-excess),  which are not present in the M-connectome at $N_F$=200K
  and $N_F=1000K$, respectively. The numbers of edges in the corresponding graph are indicated at each figure. The number of edges is inherited in the target graph at $N_F=1000K$ from the graphs at $N_F=200K$. Explicitly, the graph (c) inherits all edges from the graph (a).  The 2170 edges from the graph (b) appear in the common connectome (c), whereas 1195 edges of the graph (b) are inherited as the excess edges in the graph (d).}
\label{fig-Fexc4}
\end{figure*}

To proceed, we first identify all edges that (although with different
weights) are common for  both F-connectome and M-connectome, here
called $C^{F\&M}$-connectome at different $N_F$.  Table\ \ref{tab-EvsNf}
and Fig.\ \ref{fig-shema} summarise the number of edges and mutual relationships
of different connectomes. Fig.\ \ref{fig-Fexc4} shows the
corresponding graphs with the labelled brain regions, obtained for $N_F$=200K and 1000K.
Specifically, we find that: 
\begin{itemize}
\item The number of established edges in each considered connectome increases with the number of fibres launched $N_F$;
\item The common $C^{F\&M}$-connectome practically coincides with the M-connectome at each $N_F$, whereas the F-connectome contains an increasing number of excess edges with the increasing $N_F$;
\item The common  $C^{F\&M}$-connectome at a higher $N_F$ inherits all edges from the $C^{F\&M}$-connectome at a lower $N_F$; 
\item A significant fraction of the excess edges found in the F-connectome at a lower $N_F$ appear in the common $C^{F\&M}$-connectome but at a higher $N_F$;
\item There is a large number of the excess edges in the F-connectome
  that are never found in the common $C^{F\&M}$-connectome at a higher $N_F$; the patterns of these edges make the fundamental difference between the human female and male connectomes.
\end{itemize}

\subsection*{The Structure of Simplicial Complexes in Brain
  Graphs\label{sec:AT}}
According to Table\ \ref{tab-EvsNf} and Fig.\
\ref{fig-shema}, at each $N_F$, the common $F\&M$-connectome practically coincides with the male connectome (apart from the exact weights of edges) while there are many excess edges in the female connectome. Here, by applying $Q$-analysis (see Methods) to the corresponding graphs  at different numbers of fibres $N_F$, we show that (i) the common human connectome possesses a nontrivial hidden structure encoding multi-vertex connectivity; (ii) the excess edges of the F-connectome are not random but exhibit a highly organised structure, which thus implies a specific functionality, cf.\ Fig.\ \ref{fig-Fexc4}.

\begin{figure}[!htb]
\begin{tabular}{cc} 
\resizebox{18pc}{!}{\includegraphics{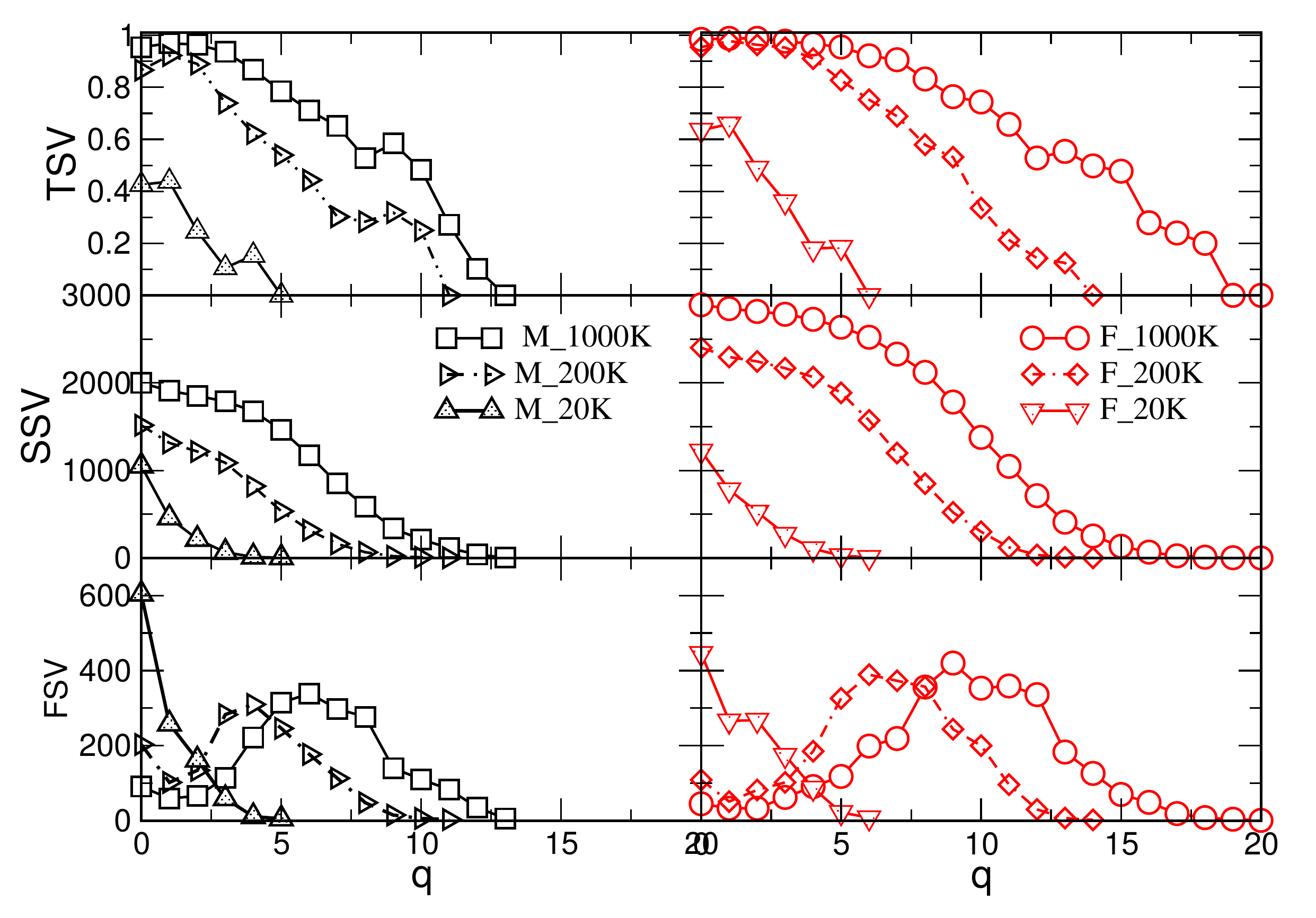}}\\
\end{tabular}
\caption{The components of three structure vectors defined in Methods
(FSV,SSV, TSV) plotted against  the topology level $q$  for
  the consensus connectomes determined from
  100 male (left) and 100 female (right) subjects 
with the varied number of fibres $N_F$, indicated in the legend.}
\label{fig-SVsx2B}
\end{figure}

In Fig.\ \ref{fig-SVsx2B} the results for the three structure vectors, defined in
Methods, are presented for different $N_F$.
As  Fig.\ \ref{fig-SVsx2B} shows, the structure of connectomes becomes
richer with the increased number of fibers $N_F$. In particular, the
cliques of a systematically larger order $q$ appear and the degree of their
inter-connectivity increases as measured by TSV. Moreover, the larger number of edges in
the F-connectome leads to a much richer structure of the simplicial
complexes, which is expressed by all structure vectors, cf. right panels
of Fig.\ \ref{fig-SVsx2B}.
We also notice that the difference between the M- and F-connectomes systematically increases with the increased $N_F$.
Representative quantitative properties are given in Table\ SI-I and Table\ SI-II in Supplementary Information.
 Noticeably, the $Q_{q=0}$ component of the FSV, which gives the number of fragments of the graph, suggests that besides the largest component some vertices and small clusters remain disconnected. The
number of fragments decreases and the connectivity increases with the
increasing $N_F$. The corresponding number of edges in the largest
cluster is given in Table\ \ref{tab-EvsNf}.  The organisation of the
present edges at each $N_F$ manifests in the presence of simplicial
complexes with the largest order $q_{max}$. From Fig.\
\ref{fig-SVsx2B} and Table\ SI-II, we see that the F-connectome possesses the cliques of a higher order; the difference increases from $q_{max}^M=5$ and $q_{max}^F=6$, at 20K, to  $q_{max}^M=13$ and $q_{max}^F=20$, at 1000K.
The number of cliques of the highest order is different, as well as
their connection to the other cliques at the level just below the
$q_{max}$.  Apart from the increased number of topology levels, the F-connectome also exhibits a significant degree of interconnections between the big cliques. 
For example, the TSV for the F-connectome at the level $q=13$, which equals to $q_{max}^M$, is still very high, about 55\%. 
Below, we identify the excess edges in the F-connectome and examine
the patterns which they make. Table\ \ref{tab-summary} shows a brief summary of different graphs' properties.

\subsection*{Hyperbolicity of the human connectome\label{sec:HB}}
Neuroanatomy of the brain enclosed in a small volume of the
skull was interpreted by the brain network which is embedded in a
hyperbolic space \cite{HB-brainSpace}. Theoretically, the
hyperbolicity of a path-connected geodesic metric space was proved
\cite{HB-geodesi-graph1991,HB-spaces-decomp2004} to be equivalent to the hyperbolicity of the graph associated with it. 
In the brain graphs studied above, the hierarchical organisation of
simplicial complexes reduces the distances between nodes in the
graph's metric space, which implies their hyperbolicity.  Here, using the 4-point Gromov criterion 
(see Methods), the hyperbolicity parameters are determined for F- and
M-connectomes obtained by varying the number of fibres $N_F$. In this context, we consider the corresponding adjacency matrix of the largest connected cluster as an unweighted symmetrical graph.  Fig.\ \ref{fig-HBx2BK} shows the results for the largest available $N_F=1000K$. In the bottom panels, the histograms of the distances between all pairs of vertices are plotted. Although the diameter  $D=8$ applies to both graphs, typical distances in the F-connectome appear to be smaller.  
In the top panels, we plot the values of the
$\delta$-parameter against the minimum distance $d_{min}$ of a given 4-tuple, as described in Methods. Specifically, lower sets of curves represent the average value $\langle \delta \rangle$ for a given $d_{min}$. Whereas the top lines contain the recorded maximum value $\delta_{max}$ from all considered 4-tuples.  

We observe that the values of $\langle \delta \rangle$ are very low,
practically never exceed 0.25, which suggests the impact of the types
of local structures populated by cliques. They are 0-hyperbolic
subgraphs (atoms) \cite{Hyperbolicity_cliqueDecomposition2017} and induced cycles, whose hyperbolicity depends on the length of
the cycle and can be expressed as a multiple of 1/4 \cite{HB-Bermudo2016}.
Moreover, $\delta_{max}=3/2$ suggests that dominant isometric subgraphs,  which determine the value of $\delta_{max}$ for the whole graph \cite{Hyperbolicity_chordality2017} in both connectomes, can be cycles $C_n$ that have $n\geq 6$ but with the diameter $D\geq 3$. 
While we regularly obtain $\delta_{max}=3/2$ in the M-connectome, it
was necessary to sample $10^9$ different 4-tuples to find it in the
F-connectome. Meanwhile, the value of $\delta_{max}=1$ occurs often in
the F-connectome. It suggests that the dominant subgraphs can be composed of cliques that are one-edge apart, which,  according to the results in
\cite{Hyperbolicity_cliqueDecomposition2017,we_SA2017}, yields that $\delta_{max}=\delta_{clique}+1$ or they contain short cycles isomorphic to 4-cycle \cite{HB-Bermudo2016}.
The situation is considerably different at the lower number of fibres where both F- and M-connectomes have gradually fewer edges (see Table\ \ref{tab-EvsNf}).
Consequently, the distances between vertices increase as well as the diameters of the graphs. The increased distances lead to the appearance of larger cycles and yield the distortion of the hyperbolicity parameter \cite{HB-distortionDelta2014} while the graphs remain hyperbolic; we find the upper bound $\delta_{max}\leq 4$ in both connectomes, as shown in Fig.\ \ref{fig-HBx2BK}.

\begin{figure}[!htb]
\begin{tabular}{cc} 
\resizebox{12pc}{!}{\includegraphics{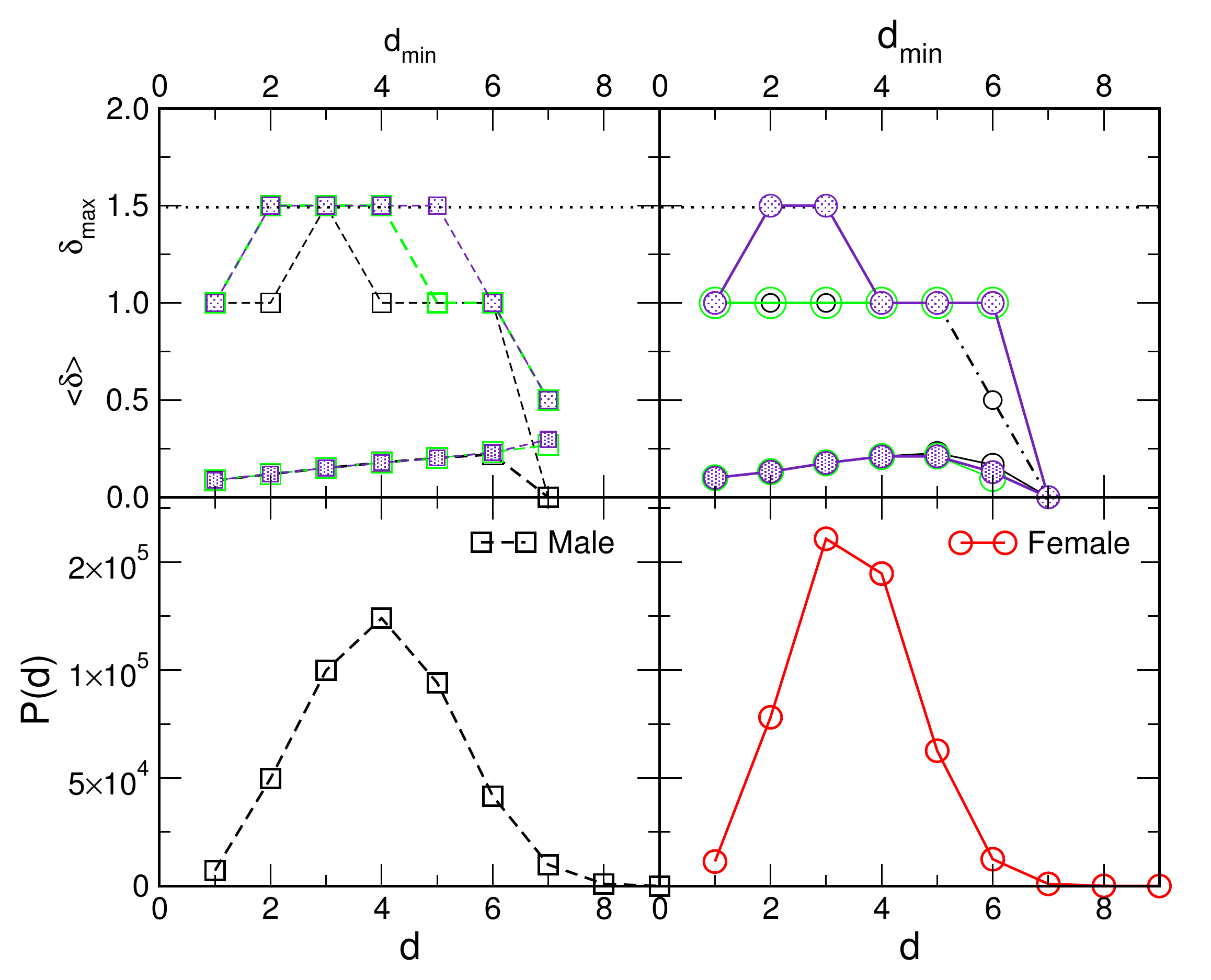}}\\
\end{tabular}
\caption{Top panles: Hyperbolicity parameters $\delta_{max}$ (upper
  curves) and $\langle \delta \rangle$ (lower curves) of the consensus connectome of female (right) and male (left) for $N_F=$1000K fibres launched. Three lines are for
$10^7$, $10^8$ and $10^9$ sampled 4-tuples. Lower panels: The distribution $P(d)$ of the shortest-path distances  $d$  for the corresponding female and male connectomes.}
\label{fig-HBx2B}
\end{figure}

\begin{figure}[!htb]
\begin{tabular}{cc} 
\resizebox{12pc}{!}{\includegraphics{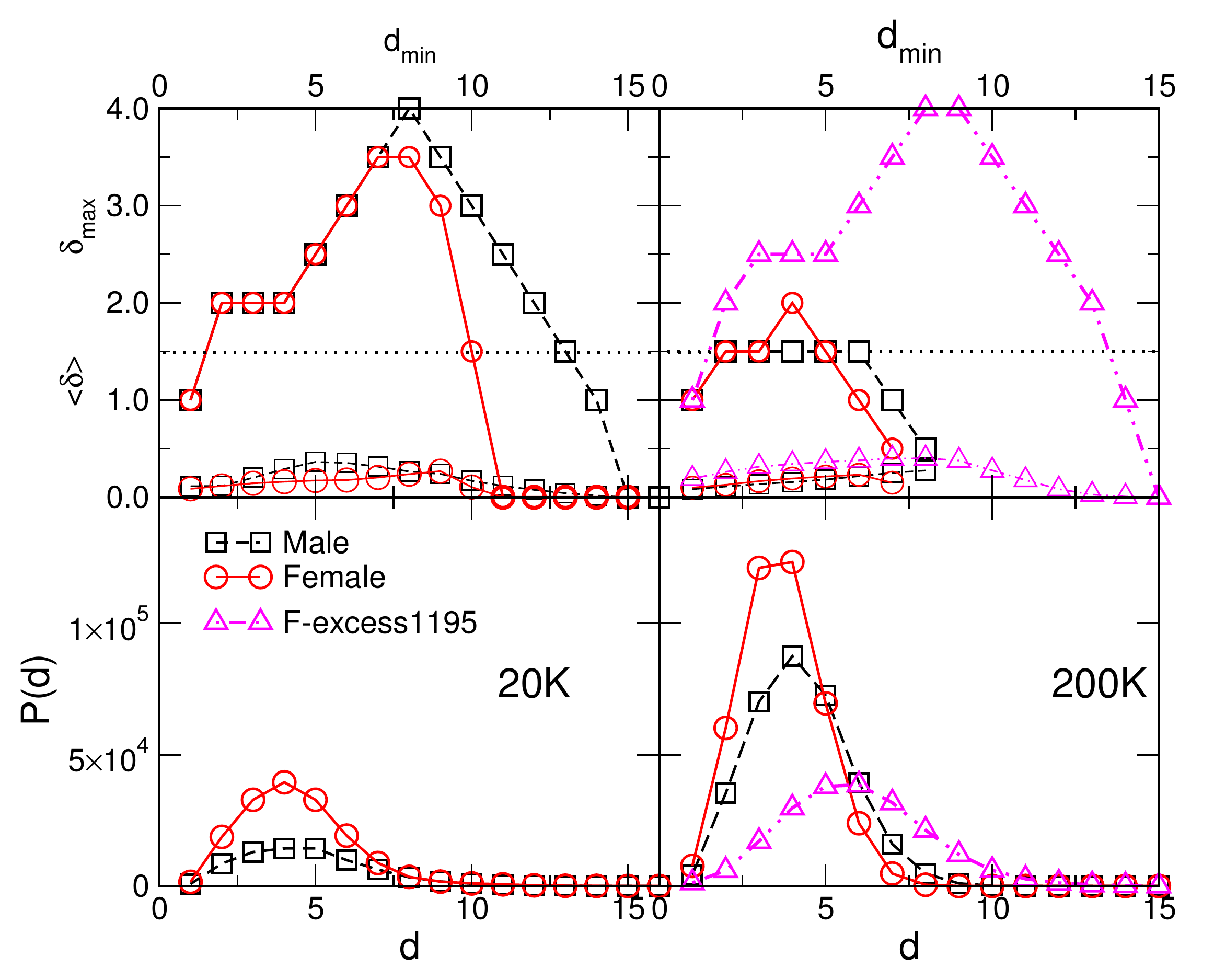}}\\
\end{tabular}
\caption{Hyperbolicity parameters  $\delta_{max}$  and $\langle \delta \rangle$   (top panels) and the shortest-path distances distribution (bottom panels) of the consensus female and male connectomes for the numbers of fibres $N_F=$ 20K (left) and 200K (right), shown at the same scale. The additional lines with triangle symbols in the right panels correspond to the excess edges in the female connectome at 200K, described in the text as F-excess1195. The number of sampled 4-tuples is $10^9$.}
\label{fig-HBx2BK}
\end{figure}

\subsection*{The structure of common F\&M-connectome and the excess edges in Female
  connectome\label{sec:excessF}}
\begin{figure*}[!htb]
\begin{tabular}{cc} 
\resizebox{24pc}{!}{\includegraphics{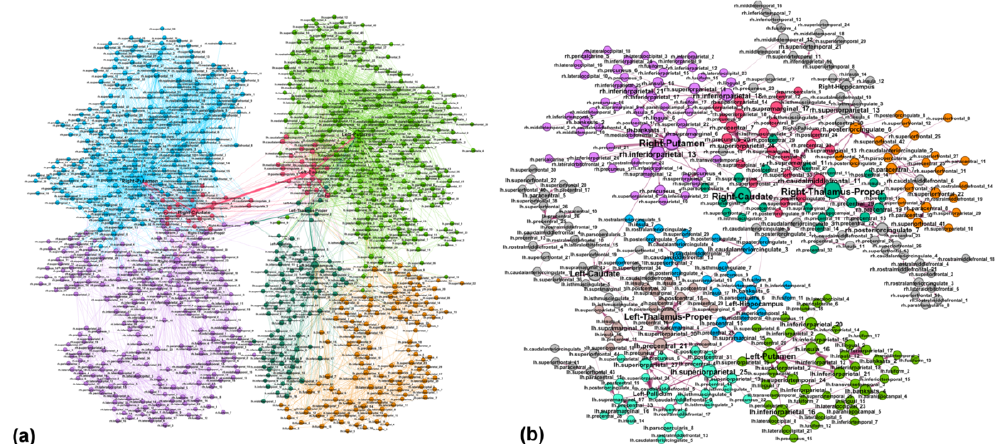}}\\
\end{tabular}
\caption{(a) The common F\&M-connectome at $N_F=1000K$ with labelled brain regions belonging to the brain anatomical communities, indicated by different colours. Weights of the edges are from the M-connectome.
(b) The robust structure of the excess connections among brain regions
(labels) in the consensus female connectome that cannot be found in
the consensus male connectome with up to 1000K fibres
launched. Different colours indicate weighted communities. We show
only the 490 edges with the significant weight in the tale of the
weight distribution, cf.\ Fig.\ SI-1, and the involved 348 brain regions.}
\label{fig-Mconn1000K-Fexcess}
\end{figure*}

By performing the edge-by-edge comparisons in the corresponding graphs, see Fig.\ \ref{fig-Fexc4}, we identify every  edge  in
terms of its source and destination vertex and the weight.
For the highest $N_F$, the common $F\&M$-connectome consists of 7110
edges which coincide with the structure of the M-connectome, cf.\
Table\ \ref{tab-EvsNf} and Fig.\ \ref{fig-shema}.   The corresponding
network of the M-connectome, as shown in Fig.\
\ref{fig-Mconn1000K-Fexcess}a, possesses a characteristic community structure related to different anatomical brain regions.  Apart from the heterogeneity of the structure due to different degrees and weights of edges, this community structure is essential for the brain functional complexity
\cite{franco2014,brainCommunities_FrontNeurosci2010,brain-mesoscopic1,brain-mesoscopic2,NatRevNeuro_segregIntegration2015,brainNets_tasksPNAS2015,Dyn-complexity-SciRep2016}
for both  F- and M-connectomes.
As mentioned above,   the F-connectome possesses an extra structure on the
top of the common F\&M-connectome; it consists of many edges that
connect different brain regions.  The number of the extra edges varies
with the number of launched fibres $N_F$, as shown in Table\
\ref{tab-EvsNf}. A subgraph of the identified excess  edges in the
F-connectome, here termed \textit{F-excess1195}, consists of 1195 edges which
systematically appear in the F-connectome,
first at $N_F=200K$ and then at $N_F=1000K$ with increased weights; these edges are not present in the corresponding M-connectomes, and thus are not part of the universal $F\&M$-connectome at the largest $N_F$. A part of this graph, containing only the edges of a substantial weight, is shown in Fig.\ \ref{fig-Mconn1000K-Fexcess}b. In the Supplementary Information list L-I, the names of source and target brain regions of these edges are given. The complete graph \textit{F-excess1195} is also shown in Fig-SI-3.

It should be stressed that the excess edges observed in the F-connectome are attached to the central brain graph, the common F\&M-connectome, at a large number of vertices. By considering F-excess as a \textit{separate graph}, cf.\ Fig.\ \ref{fig-Mconn1000K-Fexcess}b, we observe that  these  excess edges make nonrandom patterns and have a significant variation in weights (cf.\ Fig. SI-1); they  involve 348 different brain
areas in both hemispheres as well as the edges that connect the left
and right hemispheres.
The properties of the F-excess1195 subgraph are also summarised in
Table\ \ref{tab-summary}, and the distribution of distances $P(d)$, as
well as the hyperbolicity parameters with $\delta_{max}=4$ are shown in Fig.\
\ref{fig-HBx2BK}.  Noticeably, the pattern of these extra connections in the F-connectome adds some larger cycles and 112 triangles. 
However, they are well embedded in the structure of the F\&M-connectome, such that they do not appear as isomorphic cycles, and, consequently, do not increase the hyperbolicity parameter of the F-connectome. 
For comparison, we show the corresponding
features of the randomised version of the F-excess1195 graph.  
Note that for this purpose we randomise the edges \textit{within each
  hemisphere separately} while keeping the cross-hemisphere edges
intact, so that the brain anatomical structure is observed. The
parameters of the randomised graph are also shown in Table\
\ref{tab-summary}. Note that several other 
graph-theoretic properties, see the studies in reference \cite{GD_Bud_PLOS2015},
also differ in female and male connectomes.

\begin{table}[!h]
\caption{Summary of graph parameters for the F-connectome and the
  M-connectome (which is equivalent to the  common F\&M-connectome)  and the
  excess edges (F-excess) in the F-connectome at 1000K. The parameters 
  of the F-excess1195 and its
  subgraph with large weights of edges F-ex1195w18, as well as its randomised version are shown. The quantities are computed for undirected graphs: the
  average degree $<k>$, path length
  $<\ell>$ and clustering coefficient $<Cc>$, the graph's density
  $\rho$, modularity $mod$ and (the number of communities), diameter
  $D$, hyperbolicity parameter $\delta_{max}$, and the highest
  topology level $q_{max}$ with the number $(Q_{q})$ of the simplexes of that order.}
\label{tab-summary}
\begin{tabular}{|c|cc|ccc|c|cc|c|}
\hline
graph& $<k>$& $<\ell >$& $<Cc>$& $\rho$& $mod$&$D$&$\delta_{max}$&$q_{max}$\\
\hline
F-conn (Fig.1)&12.07&3.45&0.69&0.025&0.59 (6)&8&3/2&20 (1)\\
M-conn (Fig.7a) &7.01&3.97&0.67&0.014&0.62 (6)&8&3/2&13 (6)\\
\hline
F-excess (Fig.3d)&4.17&4.36&0.13&0.008&0.654&11&5/2&3 (149)\\
F-excess1195 (Fig.SI-3)&1.77&5.91&0.064&0.005&0.689&17&4&2 (112)\\
F-excess1195w18 (Fig.7b)&1.41& 6.54&0.031&0.008& 0.764&19&4&2 (18)\\
randomised-F-excess1195&0.94&9.95&0.006&0.003&0.898&30&5&2 (1)\\
\hline
\end{tabular}
\end{table}

\section*{Discussion \label{sec:discussion}}
By analysing the HCP data provided at the Budapest connectome server, we
acquired three sets of networks representing the consensus female and
male connectomes at different numbers of launched fibres 20K,
200K, and 1000K. In addition to the standard graph parameters,  by
using algebraic topology methods we
discovered a latent geometry that encodes higher-order connections in
these brain graphs. Our main findings are:
\begin{itemize}
\item \textit{Higher-order connectivity of the common F\&M-connectome.}
We have shown that the human connectome, consisting of the edges that
are common to both F\&M connectomes, possesses a hidden structure beyond
the node's pairwise connectivity. The higher-order connections between
the groups of brain regions are suitably encoded by simplexes
organised  into larger complex structures and quantified by structure vectors, cf.\ Fig.\
\ref{fig-SVsx2B}. Remarkably, the complexity of the human connectome
increases with the number of launched fibres, reaching the simplicial
complexes of the order $q_{max}+1=14$ at $N_F=1000K$. Specifically,
there are six such cliques, which contain nodes in different brain
modules (see Fig. SI-2 and the list L-I  in Supplementary Information).
We note that these simplicial complexes belong to different communities, which are anatomical mesoscopic structures of the brain graphs, cf.\ Fig.\ \ref{fig-Mconn1000K-Fexcess}a.
This architecture of connections in the brain graphs can be
characterised by the tools of hyperbolic geometry. In particular, we
find that they are Gromov hyperbolic graphs with small hyperbolicity
constant $\delta_{max}=3/2$, which characterises both F- and
M-connectomes at 1000K launched fibres.  Hyperbolicity varies with the
network density, which is directly related to $N_F$.  In contrast,
randomised (separately within each hemisphere) links exhibit much
smaller simplexes ($q_{max}^{rand}=3$) and increased hyperbolicity
parameter that points to larger cycles. These findings indicate that
the brain functional geometry consists of massive simplicial complexes
as part of anatomical communities within each hemisphere as well as cycles that connect different regions inside and between the two hemispheres.

\item \textit{Structure of the excess edges in F-connectomes.}
F-connectome systematically appears to be better connected, i.e., has
a more significant number of edges at every $N_F$. Here, a more detailed
inspection of the source-and-target brain region and the weight that
identifies an edge indicates that two groups of excess edges occur:
(1) The edges appearing in the F-connectome at  a relatively low
number of fibres which can appear in the M-connectome but only if a
much larger number of fibres is launched;  (2) The edges that robustly
appear only in the F-connectome and have not been established in the
M-connectome, including the highest available number 1000K of fibres.
From the second group, the identity of 1195 edges that first appear at
200K in the F-excess subgraph and are not present in the common F\&M-connectome at 1000K are given in Supplementary Information. In particular, Fig.SI-3  shows the complete graph, while the list L-II contains only the edges with large weights. A comparison with the (inside the hemisphere) randomised graph has shown that these F-excess edges, considered as a separate graph, also have an organised structure involving a large number of brain regions, cf.\ Fig.\ \ref{fig-Mconn1000K-Fexcess}b. Direct analysis and its hyperbolicity parameter suggest a geometry dominated by cycles and small simplexes.
\end{itemize}

To summarise, our study reveals how the \textit{functional geometry} of human connectome can be expressed by higher-order connectivity, described by simplicial complexes and
induced cycles. This kind of structure is built into the anatomical
communities of the brain at the mesoscopic scale in both brain
hemispheres. However, the precise role of these simplicial complexes
for the dynamical segregation in brain functional complexity  remains
to be better understood.   In this context, the developed methodology provides
new topological measures of the consensus brain networks and
quantifies the robust gender differences. Specifically, a part of
connections is more natural to invoke in the female than in the male
brain, where much more fibres need to be launched to identify
them. Whereas the other fraction of such connections consists of edges that appear exclusively in the consensus female
connectome, they have not been identified in the consensus male
connectome.

It should be stressed that the considered consensus networks represent
a kind of typical structures with the fixed number of vertices as 1015
brain regions while the edges are common for all 100 male and
similarly for all 100 female, recorded within HCP in a representative
set of (young and healthy) individuals.  
 Note that, in each particular subject, the number of brain
 connections can deviate, e.g.,  being even considerably more abundant
 than in the respective consensus network. Moreover, the structure of
 possible connections is expected to vary with age, particular practice and with a development of diseases. 
Based on the brain imaging data, the methodology developed  in this work
would be suitable to reveal subtle differences between pairs of brains
as well as changes in the brain of the same individual.  Similar
studies have been done with the patterns induced by the brain
spontaneous fluctuations and content-related activity recorded by EEG
\cite{we-MBNets-PLOS2016,we-Frontiers2018,BrainspontaneousFluct_NatRev2007},
complementing the traditional methods. The application of our
methodology to these issues warants a
separate study which would include  a more detailed investigation of
the role of orientation and the weights of the edges.

\section*{Conclusions\label{sec;conclusions}}
Our analysis has revealed that the human connectome possesses a
hyperbolic geometry and a complex structure on the scale between the
node's edges and the mesoscopic anatomical communities within  the
cerebral hemispheres. This structure, composed of simplicial complexes
of different sizes and cycles that connect them, accurately describes
the higher-order connectivity among different regions of the brain,
divided into anatomical modules. Therefore, it can provide a reliable
basis for understanding the functional complexity of the
brain. Moreover, the female connectome appears to have a structure
different from the common F\&M-connectome, not only in the number of
edges but also in its  organisation expressed by these higher-order
connections.  It might be conjectured that these excess connections
imply additional functionality of the female connectome, which can
have evolutionary, biological, biochemical, and even social
origins. These issues go beyond our mathematical analysis of brain
graphs. However, we believe that our  findings can  motivate further
studies to better understand the origin and functional consequences of the apparent gender differences in the human connectome.


\section*{\normalsize Acknowledgments}
Work supported  by  the Slovenian
Research Agency (research code funding number P1-0044). MA received
financial support from the Ministry of Education, Science and
Technological Development of the Republic of Serbia, under the project
OI 174014.  RM is also grateful for the NSERC and CRC programs for
their support.

\end{document}